\input harvmac
\def\abstract#1{
\vskip .5in\vfil\centerline
{\bf Abstract}\penalty1000
{{\smallskip\ifx\answ\bigans\leftskip 1pc \rightskip 1pc 
\else\leftskip 1pc \rightskip 1pc\fi
\noindent \abstractfont  \baselineskip=12pt
{#1} \smallskip}}
\penalty-1000}
\def\fc#1#2{{#1\over #2}}
\def\frac#1#2{{#1\over #2}}

\def\br{\hfill\break}
\def\ni{\noindent}

\def\mark{}
\def\al{\alpha}
\def\eps{\epsilon}

\def\p{\partial}

\def\bx#1{{\bf #1}}
\def\cx#1{{\cal #1}}
\def\tx#1{{\tilde{#1}}}
\def\hx#1{{\hat{#1}}}

\def\us#1{\underline{#1}}
\def\hth/#1#2#3#4#5#6#7{{\tt hep-th/#1#2#3#4#5#6#7}}
\def\nup#1({Nucl.\ Phys.\ $\us {B#1}$\ (}
\def\plt#1({Phys.\ Lett.\ $\us  {B#1}$\ (}
\def\cmp#1({Comm.\ Math.\ Phys.\ $\us  {#1}$\ (}
\def\prp#1({Phys.\ Rep.\ $\us  {#1}$\ (}
\def\prl#1({Phys.\ Rev.\ Lett.\ $\us  {#1}$\ (}
\def\prv#1({Phys.\ Rev.\ $\us  {#1}$\ (}
\def\mpl#1({Mod.\ Phys.\ Let.\ $\us  {A#1}$\ (}
\def\atmp#1({Adv.\ Theor.\ Math.\ Phys.\ $\us  {#1}$\ (}
\def\ijmp#1({Int.\ J.\ Mod.\ Phys.\ $\us{A#1}$\ (}

\def\subsubsec#1{\ \br \noindent {\it #1} \br}
%
\input epsf
\noblackbox
%
\def\eqsp{\phantom{\pmatrix{\big(\big)\cr\big(\big)}}}
\def\ca{c}
\def\bal{\bb \al}
\def\vtv{V_{TV}}
\def\const{{\scriptstyle \rm const.\ }}
\def\mt{\tilde{m}}
\def\mm{{\Omega}}
\def\msusy{M_{susy}}\def\vk3{V_{K3}}
\def\kt{\cx K}

\def\mstr{M_{str}}
\def\cc{c.c.}\def\hc{h.c.}
\def\bb#1{{\bar{#1}}}

\def\Si{\Sigma}
\def\La{\Lambda}
\def\be{\beta}

\def\Im{{\rm Im}\, }
\def\rk{{\rm rk}\,}
\def\ss{\scriptstyle}
\def\pp#1#2{P^#2_#1}
\def\vp{\langle {\rm Tr}\, \phi^2 \rangle}
\def\qs#1{e^{#1 \pi i s}}
\def\qt#1{e^{#1 \pi i t}}
\def\qt#1{q_t^#1}

\lref\SWi{N. Seiberg and  E. Witten, \nup {426} (1994) 19,  
        erratum: ibid  {430} (1994) 396.}%
\vskip-2cm
\Title{\vbox{
\rightline{\vbox{\baselineskip12pt\hbox{CERN-TH/2000-083}
                                  \hbox{hep-th/0003198}}}}}
{On Supersymmetry Breaking in String Theory} \vskip-1cm \centerline{\titlefont and its Realization in Brane Worlds}

\abstractfont 

\vskip 0.8cm
\centerline{P. Mayr}
\vskip 0.6cm
\centerline{CERN Theory Division} 
\centerline{CH-1211 Geneva 23}
\centerline{Switzerland}
\vskip 0.3cm
\abstract{%
We use string duality to describe
instanton induced spontaneous supersymmetry breaking 
in string compactifications with additional background fields. 
Dynamical supersymmetry breaking by space-time
instantons in the heterotic string theory
is mapped to a tree level breaking in the type II 
string which can be explicitly calculated 
by geometric methods. 
The point particle limit describes the non-perturbative
scalar potential of a SYM theory localized on a 
hypersurface of space-time. The $\cx N=0$ vacuum displays condensation 
of magnetic monopoles and confinement. The supersymmetry breaking
scale is determined by $\mstr$, which can be 
in the TeV range, and the geometry transverse to the gauge theory.
}

\Date{\vbox{\hbox{ {March 2000}}
}}
\goodbreak

\parskip=4pt plus 15pt minus 1pt
\baselineskip=15pt plus 2pt minus 1pt
\leftskip=8pt \rightskip=10pt
%

\newsec{Introduction and summary}
The remarkable success in the understanding of non-perturbative 
properties of string theories triggered by string dualities can not 
conceal the fact that their impact on the formulation of  
theories with $N<2$ supersymmetry and dynamical supersymmetry 
breaking has been much less effective. In fact 
the step to $N=1$ supersymmetry may be 
already a vital one for the description of the non-supersymmetric world as
it is known that dynamical supersymmetry breaking may appear in these
theories as a 
consequence of strong gauge interactions
\ref\gc{H.P. Nilles, \plt 115 (1982) 193, \nup 217 (1983) 366;\br
S. Ferrara, L. Girardello and H.P. Nilles, \plt 125 (1983) 457;\br
for a review and further references see H.P. Nilles, \ijmp 5 (1990) 4199.}.
It is interesting
that supersymmetry breaking in the string theory might be so
closely related to its low energy sector. In fact in the realization of
confinement as result of monopole condensation 
\ref\Hoo{G. t'Hooft, \nup 190 (1981) 455.}%
\SWi, anomaly
considerations imply a non-zero gaugino condensate 
\ref\Kon{K. Konishi, \plt 392 (1997) 101.}
which 
breaks supersymmetry in the theory coupled to gravity \gc.
In this sense the lack of supersymmetry observed in our world could be 
linked in an intriguing way to the existence of confining gauge theories.

One of the most useful $N=2$ dualities has been the one between
type II and heterotic strings 
\ref\KV{S. Kachru and C. Vafa, \nup {450} (1995) 69.}%
\ref\FHSV{S. Ferrara, J. A. Harvey, A. Strominger
                  and C. Vafa, \plt {361} (1995) 59.}.
As the heterotic coupling
maps to a geometric modulus in the type II theory, non-perturbative
effects of the heterotic theory are related to tree level of the 
type II string. In particular space-time instanton effects map
to geometric instantons in the type II theory which can be 
calculated by mirror symmetry. More fundamentally the instantons
in the type II theory can be viewed as honest space-time instantons
in the RR-sector, which has the special property to not depend on
the type II string coupling and thus is described by string tree level.

This opens the fascinating possibility that after a supersymmetry
breakdown to $\cx N=1$, dynamical supersymmetry breaking to 
$\cx N=0$ in the RR-sector of the type II theory is calculable 
as the non-perturbative effects are still governed
by a geometric coupling constant and not the string coupling.
In particular condensation of fermions will appear at string tree level.
This non-supersymmetric type II theory will be dual to a $\cx N=1$
supersymmetric heterotic theory where supersymmetry is
broken only in the non-perturbative sector of the string theory.
Such a picture was advocated in 
\ref\Kasi{S. Kachru and  E. Silverstein,
\nup 463 (1996) 369.}
using the special constructions 
of \ref\Rvwad{C. Vafa and E. Witten, 
        Nucl. Phys. Proc. Suppl. $\us {46}$ (1996) 225.}
to obtain $\cx N=1$ dual pairs from $\cx N=2$ ones 
by freely acting 
orbifolds\foot{The relation between spontaneous supersymmetry
breaking in string theory 
and freely acting orbifolds with various supersymmetries
has been studied in refs.%
\ref\Rkk{E. Kiritsis and C. Kounnas, \nup 503 (1997), 117;\br
E. Kiritsis, C. Kounnas, P. M. Petropoulos and J. Rizos, \nup 540 (1999) 87.}%
.}%
. Based on qualitative arguments on the general properties
of the orbifolding procedure it was argued that indeed 
gaugino condensation in the heterotic theory should map to 
a tree level supersymmetry breaking in the type II theory which,
remarkably enough, was related to monopole condensation and
the associated relalization of confinement. Unfortunately 
the orbifold duals of \Kasi\ are not accessible to instanton calculations by
mirror symmetry and thus the study of string duals
with supersymmetry breaking has been restricted to a rather
qualitative level, far from the success in the $\cx N=2$ supersymmetric
case.
To provide a calculable framework of string duals with dynamical
supersymmetry breaking is one of the goals of this paper.

In a seemingly unrelated development in the supergravity part of the world,
the no-go theorem 
 \ref\CGP{S. Cecotti, L. Girardello and M. Porrati, 
\plt 145 (1984) 61.}
on partial local supersymmetry breaking $N=2\to N=1$ 
was deprived its validity 
\ref\FGP{S. Ferrara, L. Girardello and M. Porrati, \plt 366 (1996) 155;
\plt 376 (1996) 275.}. 
The argument is that for a 
special class of supergravity theories, the tensor calculus of 
ref. 
\ref\LPW{B. de Wit, P. Lauwers and A. Van Proeyen, \nup {255} (1985) 269.}%
, which was the basis for the derivation of the no-go theorem, 
is inappropriate. In these special cases, the no-go theorem does not 
apply and spontaneous partial supersymmetry breaking has indeed be shown 
to be possible \FGP.
In string theory compactifications, scalar potentials that 
may trigger spontaneous supersymmetry breaking can be induced by 
RR background fields in the type II string 
\ref\Rps{J. Polchinski and A. Strominger,
\plt 388 (1996) 736.}
or magnetic backgrounds in the dual heterotic theory 
\ref\Rhs{C. Bachas, {\it A Way to break supersymmetry}, hep-th/9503030;\br
I. Antoniadis, E. Gava, K.S. Narain and T.R. Taylor,
\nup 511 (1998) 611.}.
However, as we will argue in this paper, 
partial breaking fails to exist, although in an interesting way. 
The reason is that even though the conditions of partial supersymmetry 
breaking may be satisfied in the classical string effective theory, 
there will always be instanton effects in compactifications with 
{\it finite volume}, that restore the foundations of the no-go
theorem. We will also argue that the effect of these instantons is 
to break supersymmetry completely, rather then to restore $\cx N=2$.

An interesting picture begins to emerge by observing that the classical
string effective theories associated to $N=2$ type II/heterotic duals
fall precisely into the class of supergravities that satisfy the
necessary conditions for partial supersymmetry breaking. 
Would it not be for the instanton effects, we would
get $\cx N=1$ supersymmetric dual pairs in this way. Turning to the 
instanton corrections one observes that they
represent non-perturbative space-time effects in the heterotic string
and tree level geometric instantons in the type II theory. 
In other words, {\it the instanton effects that restore the no-go theorem
embody precisely the sought for description of non-perturbative
dynamical supersymmetry breaking}. Using mirror symmetry
in the tree level type II theory to determine the geometric instantons 
we are thus in the remarkable situation to be able to systematically
calculate the instanton effects of a phenomenologically relevant
non-perturbative supersymmetry breaking.

Interestingly, and in full agreement with the observations
made in the free orbifold models of \Kasi, 
the supersymmetry breaking is intimately related
to strong gauge interactions in a SYM theory embedded into
the string theory. In fact this theory 
is of ``the world on a brane'' type, with the 
gauge fields localized on a hypersurface of ten-dimensional
space-time. These theories have been very successfully studied 
in the $\cx N=2$ context in terms of 
curved brane geometries in string theory, 
either in the geometric engineering of 
type II strings on singularities of Calabi--Yau 3-folds 
\ref\KLMVW{A. Klemm, W. Lerche, P. Mayr, C. Vafa, N. Warner,
                \nup {477} (1996) 746.}%
\ref\KKV{S. Katz, A. Klemm and C. Vafa, \nup {497} (1997) 173.}%
\ref\KMV{S. Katz, P. Mayr and C. Vafa, \atmp 1 (1998) 53.}
or, for simple enough gauge group $G$, 
in terms of the T-dual type IIA 5-brane \KLMVW. The latter 
realization can also be reached starting from field theories
living on the world-volume of ``flat branes'' which are bent
by quantum corrections 
in a way  determined by the non-perturbative duality to 
M-theory \ref\WITFB{E. Witten, \nup 500 (1997) 3.}. 
Both, the ``flat'' and the curved brane constructions, 
give a string theory realization of the intensively discussed extra
large dimension scenarios 
\ref\Radd{N. Arkani-Hamed, S. Dimopoulos 
and G. Dvali, \plt 429 (1998) 263;\br
I. Antoniadis, N. Arkani-Hamed, S. Dimopoulos 
and G. Dvali, \plt 436 (1998) 257.}%
. In particular the ``flat branes''
have become a most popular tool in qualitative 
phenomenological studies due to their conceptional simplicity 
as compared to the geometric approach. 

So finally, taking the point particle limit 
of the eventually non-supersymmetric dual pair, leads us 
right to an extension of geometric engineering\foot{A different
approach uses F-theory compactifications
\ref\fth{
S. Katz and C. Vafa, \nup 497 (1997) 196;\br
M. Bershadsky, A. Johansen, T. Pantev, V. Sadov and
C. Vafa, \nup 505 (1997) 153;\br
P. Mayr and P. Berglund, \atmp 2 (1999) 1307.}.}
to more realistic SYM theories on the brane, with broken supersymmetry and 
patterns of confinement and chiral symmetry breaking\foot{In 
the following we will
use the definiton of a brane in the general sense of specifying a
 hypersurface on which the gauge fields localize. This notion 
is independent of the specific string theory 
construction in terms of dual D-branes, M-branes or geometrically 
engineered theories at a singularity.}. Using
the results from mirror symmetry in the type II theory 
allows to derive the explicit non-perturbative scalar potential 
in the SYM theory on the brane.

It is worth noting that 
the description of supersymmetry breaking in the SYM theory on the brane
is quite relevant for the large dimension scenarios 
one of whose primary motivations has been the hierarchy problem. 
A serious attempt to explain the apparent weakness of gravity 
should be based on a framework of quantum gravity such as string theory,
and not just field theories on ``flat branes''.
Note that the resurrection of the no-go theorem implies that 
the complete breakdown to $\cx N=0$ (rather than $\cx N=1$)
is inevitable when the SYM  is coupled to string theory.
It would be quite interesting to develop the understanding of 
the field theories based ``flat branes'' with $\cx N=1$
supersymmetry 
\ref\Rfb{J.L.F. Barb\'on, \plt 402 (1997) 59.}
to a level where the supersymmetry breakdown
can be seen, at least qualitatively. The reason that the
flat branes miss this effect is apparently the 
lack of a satisfying description of the coupling of the
world volume theory to gravity and string theory.

\mark
Note that we argued that the restoration of the no-go theorem relies on stringy
instantons on a {\it finite volume} manifold. This does not exclude
partial supersymmetry breaking in diverse non-compact limits.
In particular, taking all, transverse as well as tangential
dimensions of infinite volume gives a pure 
field theory decoupled from strings and gravity.
It should be clear that neither there is
a no-go theorem in pure field theory nor do we claim
so\foot{E.g., gaugino condensation does
{\it not} break supersymmetry in field theory
\ref\nosusybr{
H.P. Nilles, \plt 112 (1982) 455;\br
G. Veneziano and S. Yankielowicz, \plt 113 (1982) 231.}.}.
In fact, while we completed this work, 
an interesting paper appeared
\ref\TV{T.R. Taylor and C. Vafa, {\it RR flux on Calabi-Yau and 
partial supersymmetry breaking}, hep-th/9912152.},
which also describes supersymmetry breaking 
by RR fluxes in the context of geometric engineering. 
In ref.\TV\ the limit of infinite string mass $M_{str}$ 
is taken which results in a pure field theory 
decoupled from string theory where partial breaking is possible.
Moreover this reference contains an elegant derivation of how the relevant
$\cx N=1$ superpotentials follow from soliton masses in 
the presence of RR-potentials. 
\mark

This paper is organized in two parts. In the first part we consider 
the issue of partial supersymmetry breaking in the string 
effective supergravity. In section 2 we argue
that for string compactifications on manifolds with
finite volume, any classical partial breaking to $\cx N=1$ 
is followed by a further breakdown to $\cx N=0$ by instanton
effects. We resurrect the general no-go theorem on partial 
breaking for string effective theories. 
In section 3 we relate these results to the class of string effective 
theories which reduce to ordinary Super-Yang-Mills theories in 
the point particle limit. The instantons that are 
relevant for the supersymmetry breakdown turn out to be non-perturbative 
in the field theory coupling. In section 4 we describe in more detail
the mechanism of supersymmetry breaking and 
the calculation of the non-perturbative gravitino masses 
by mirror symmetry. Moreover we observe a general link 
between the breakdown of supersymmetry and that of conformal invariance.

In the second part, which starts with section 5 and can be read 
almost independently, we focus further on the pattern of supersymmetry
breaking in the point particle limit, namely in the SYM 
on the brane embedded in the string theory. The general non-perturbative 
scalar potential contains apart from the $\cx N=1$ adjoint mass term 
extra soft breaking terms that arise from the coupling 
to the string sector. We use special geometry to proof that these 
terms are mandatory and there is no partial breaking to $\cx N=1$
in the field theory coupled to string theory.
The scale of supersymmetry breaking terms is determined by the string scale
$\mstr$, which is argued to be favorably in the TeV range, 
multiplied by a factor
$\sim b_1\cdot \vtv^{-1}$, where $b_1$ is the one-loop 
beta function coefficient of the SYM theory 
and $\vtv$ the volume of the dimensions
transverse to the gauge theory.
Details on the supergravity calculations are relegated
to the appendices.

\newsec{Partial supersymmetry breaking in $\cx N=2$ supergravity
 \hskip 17.1291cm  a ({\it  or ``Two into one still won't go''}) }

Before focusing on the string theory embeddings of SYM theories
let us ask in general what patterns of supersymmetry
breaking may appear in the $\cx N=2$ string effective theory.
As far as partial local supersymmetry breaking by a Super-Higgs effect
is concerned, there used to be a very strong result, the ``Two into one won't
go'' theorem of \CGP. It states that a zero eigenvalue of the 
gravitino mass matrix implies that the second one is zero, too.

However, the no-go theorem is based on the existence 
of a holomorphic prepotential
$\cx F(z_i)$ that defines the effective $\cx N=2$ supergravity 
action of the $n_V$ vector multiplets. At that time this was indeed 
the only known description of the $\cx N=2$ supergravity theory \LPW.
Since then, an alternative definition  has been formulated 
which is not based on a prepotential 
\ref\CAFP{A. Ceresole, R. D'Auria, S. Ferrara and A. Van Proeyen, 
\nup 444 (1995) 92.}. It thus appeared that the 
absence of a prepotential might possibly allow
to evade the no-go theorem and indeed it was shown in \FGP\
that partial supersymmetry breaking is possible in this special 
situation. The globally supersymmetric version of this
partial breaking has been described independently in 
\ref\APT{I. Antoniadis, H. Partouche and T. R. Taylor,
\plt 372 (1996) 83.}.

The necessary condition for the absence of a prepotential and 
thus partial supersymmetry breaking is the following one. The 
framework of \CAFP\ starts from a section $\Pi =(X^\Si,F_\Si)^T$ 
of a $Sp(2n_V+2,\bx Z)$ bundle over $M_V$. It is 
invariant under symplectic transformations acting on $\Pi$:
\eqn\sptf{\Pi \to M\, \Pi, \qquad M\in Sp(2n+2,\bx Z).}
The components $X^\Si,F_\Si$ of the section $\Pi$, also called periods,
depend on the $n_V$ scalar components $z_i$ of the vector multiplets 
which parametrize an $n_V$ dimensional special K\"ahler manifold
$\cx M_V$. In a generic situation the upper components $X^\Si$ can be
thought of as homogeneous coordinates on $\cx M_V$, while
the lower half $F_\Si$ of $\Pi$ is related to a prepotential $\cx F(X^\La)$
of homogeneous degree two by $F_\Si=\p \cx F/\p X^\Si$. 
The transition to the generic inhomogeneous variables $z_i$ on 
$M_V$ is described by the matrix 
\eqn\jac{ 
A^{\Si i}=\fc{\p X^\Si}{\p z_i}.
}
For a special form of the prepotential $\cx F$ and
a special choice of the section $\Pi$, the matrix 
$A^{\Si i}$ may be degenerate. Then the $X^\Si$ can 
not serve as homogeneous coordinates on $\cx M_V$ and no 
prepotential $\cx F(X^\La)$ exists.
Note that this statement is not invariant under $Sp(2n+2,\bx Z)$
transformations and by choosing a different section $\tx \Pi=M\Pi$
one may always transform to homogeneous coordinates $\tx X^{\Si}$ in 
which a prepotential $\tx {\cal F}(\tx X^\Si)$ exists.

A string effective supergravity related to the geometric 
type II compactification on a
Calabi--Yau manifold $M$ is classically described by a cubic prepotential
\eqn\cypp{
\cx F=\fc{1}{3!} \, C_{ijk}X^iX^jX^k/X^0=(X^0)^2\,\fc{1}{3!} \,  C_{ijk}t_it_jt_k,}
where $t_i=X^i/X^0$ are so-called special coordinates that parametrize
volumes of homology 2-cycles in $M$. Depending on the intersection
matrix $C_{ijk}$ of $M$ it may be possible to choose a section $\Pi$
where the $X^\Si$ are dependent, the matrix $A^{\Si i}$  is 
degenerate and partial breaking to $\cx N=1 $ supersymmetry 
is possible%
\foot{We will argue that this happens for string theory
embeddings of SYM theories in the next section.}.

However in a finite volume there are stringy instanton corrections
to the classical prepotential \cypp\ that change this picture.
The exact instanton corrected prepotential can be 
determined \ref\Rcan{P. Candelas, X.C. De La Ossa, P.S. Green and L. Parkes,
\nup 359 (1991) 21.} using mirror symmetry\foot{For
an overview and references, see 
\ref\ms{{\it Essays on Mirror Manifolds}, (ed. S.T. Yau), 
Int. Press, Hong Kong, 1992.}. For a recent proof 
in the supersymmetric sigma model, see \ref\HV{
K. Hori and C. Vafa,
{\it Mirror symmetry}, hep-th/0002222.}.}. 
Specifically, the
components of $\Pi$ are identified under mirror symmetry with the period
integrals of the holomorphic 3-form $\Omega^{3,0}$ on the
mirror manifold $W$ of $M$:
\eqn\perint{
X^\Si=\int_{A^\Sigma}\, \Omega^{(3,0)}(W),
\qquad F_\Si=\int_{B_\Sigma}\, \Omega^{(3,0)}(W).}
Here $\{A^\Si\}$ denotes an integral basis of homology 3-cycles
and $\{B_\Si\}$ their duals. The period integrals $F_\Si$ have 
the large $t_i$ expansion
\eqn\piexp{
F_\Si=\fc{1}{2!}\, C_{ijk}t_jt_k+f(t_n,q_n),}
where $q^n=exp(2\pi i t_n)$ are the instanton corrections
suppressed by the exponential of the K\"ahler volumes  of 2-spheres.

We want to argue now that even in a situation where the 
geometrical data $C_{ijk}$ allow for a partial supersymmetry 
breaking to $\cx N=1$, the stringy instanton corrections always
lead to a complete breakdown of supersymmetry. To do so
we have to assure that after the inclusion of the instanton corrections, 
the upper part $\hx X^\Si$ of any section $\hx \Pi$ gives good homogeneous 
coordinates on $\cx M_V$. By the relation \perint\ this 
is the same as asking whether for any choice of 
a basis of $A$-cycles in \perint, the period integrals
represent locally good homogeneous coordinates on $\cx M_V$. 
Luckily this type of question is a prominent one in 
algebraic geometry and is part of the so-called infinitesimal 
Torelli problem. In the present context it has been shown in ref.%
\ref\BG{R.L. Bryant and P.A. Griffiths, {\it Some
observations on the infinitesimal period relations for regular
threefolds with trivial canonical bundle}, in  Arithmetic and
Geometry Volume II (M. Artin and J. Tate, eds.), Progress in
Math.  39, Birkh\"auser, Boston-Basel-Berlin, 1983.} that the
period integrals indeed have the required property.

In conclusion, although partial supersymmetry breaking may appear
to be possible in the classical string theory, there will be always
stringy instanton corrections that lead to a complete breakdown of 
supersymmetry  and the mechanism of refs.\FGP\ 
can not be realized in the string effective theory.

\newsec{Localized SYM theories and their string theory embedding}

Form the point of physics,  
the situation that the classical section $\Pi$  allows 
for partial supersymmetry breaking while the instanton corrected
one does not is quite a remarkable one. Specifically the
supersymmetry breaking scale will depend non-perturbatively on the
K\"ahler moduli. We will argue now that this is the case 
in the class of string effective theories realizing SYM theories
localized on a six-dimensional hypersurface of space-time.
As we will discuss in the following, the relevant modulus is 
identical to the field theory coupling constant and the
supersymmetry breakdown is thus interpreted as a non-perturbative
breaking by {\it space-time} instantons. 

\subsec{Geometry and couplings}
The geometric engineering approach \KLMVW\KKV\KMV\
starts from a type II string compactified on an almost singular
Calabi--Yau manifold $X$\foot{For a basic introduction and further 
references, see \ref\PMNELEC{P. Mayr, Fortsch. Phys. $\us{47}$ (1998) 39.}.}.
It is useful to think about this compactification
as a two step process of a compactification to six dimensions
on a singular two complex-dimensional manifold $Y$ followed by 
a further compactification on the base sphere $B$ to four dimensions.
At the singular point $p$ of $Y$ there are $r=\rk G$ small 2-spheres 
$C_i$ with an intersection matrix equal to the negative of the 
Cartan matrix $C_{ij}^G$. D2-branes wrapped on the 2-spheres $C_i$
represent $\cx N=2$ vector multiplets in six dimensions with 
a gauge kinetic term (in  a tree level approximation)
\eqn\sixdkin{
\cx L_{kin}= \fc{1}{4}\, C_{ij} F^{(i)\mu\nu}F^{(j)}_{\mu\nu}.}
In particular the effective action is independent of the 
directions transverse to the singularity, as is
expected from the fact that the gauge fields are localized 
on the six-dimensional hypersurface specified by $p\subset Y$.
This gives a natural string theory realization of the
(possibly large) extra dimension scenario of \Radd.

Upon compactification on the sphere $B$ one obtains an 
$\cx N=2$ supersymmetric QFT with gauge group $G$ in four dimensions. 
Dimensional reduction implies that
the four-dimensional tree level gauge coupling 
is given by the (complexified) K\"ahler volume $V_B$ of B:
\eqn\fdcoup{
\fc{\theta_{FT}}{2\pi}+\fc{4\pi i}{g_{FT}^2}=\int_B\, B_{NS} +i\,  
V_B\equiv s.}
The real part of $s$ describes the integral of the Neveu-Schwarz 
$B$-field on the base 2-sphere $B$.

At this point it seems useful to fix the notation for the
coupling constants. In the following we
will refer to the field theory coupling constant $s$ as 
``the coupling constant'' and use ``perturbative'' with 
respect to the the field theory which is designed to describe our world. 
The string theory embedding with $s$ identical to the string coupling 
constant is
a heterotic string on K3$\times T^2$. It is dual
\KV\FHSV\
to the geometric type II compactification where $s$ is the geometric volume
\fdcoup\ and {\it not} the string coupling. This is the usual story used
in geometric engineering: we can calculate non-perturbative phenomena
of the effective {\it field} theory by using the geometric type II picture. 
In the present paper we extend this correspondence to also 
calculate non-perturbative quantities of the string effective
{\it supergravity} theory to which the SYM theory is coupled.

\subsec{The string effective supergravity}
The four-dimensional tree level coupling \fdcoup\ fixes the 
classical piece of the prepotential to be
\eqn\clpp{
\cx F = -s\, C_{ij}t_i t_j+\dots,}
where the dots denote $\al'$ and finite coupling corrections.
It is easy to see that the string effective supergravity 
obtained from geometric engineering has precisely the property 
to allow for a partial supersymmetry breaking. In the 
inhomogenous coordinates $t_i,\ s$, the standard section 
$\Pi=(X^\Si,\p \cx F/\p X^\Si)$ is
\eqn\standsec{
X^\Si=(1,s,t_i),\qquad 
F_\Si=(2\cx F-s\, \cx F_s-\sum_{k}t_k\, \cx F_k,\cx F_s,\cx F_i),
}
where subscripts denote differentiation. From the prepotential 
\clpp\ one obtains after a simple symplectic transformation 
$(s,\cx F_s)\to (\cx F_s, -s)$ a section $\tx\Pi$ with 
\eqn\clpv{
\tx X^\Si=\pmatrix{1,\cx F_s,t_i}^T,\qquad 
\tx F_\Si=\pmatrix{2\cx F-s\cx F_s-\sum_k t_k \cx F_{t_k},-s,\cx F_{i}}^T,}
In particular
\eqn\sht{
\cx F_s=\p\cx F/\p s =-C_{kl}t_k t_l.}
Note that the $\tx X^\Si$ do not depend on $s$ and thus the necessary 
condition for partial supersymmetry breaking is satisfied.

However, as we argued on general grounds in section 2,
instanton corrections will change this picture. In the present situation
the relevant instantons have an action $\sim e^{2\pi i s}$ 
which is characteristic of space-time instantons as interpreted in the
field theory. It is one of the most powerful aspects of the 
geometrically engineered quantum field theories (GEQFT's) 
that one may determine these non-perturbative corrections
from the dual, geometric type II instantons by mirror symmetry. 
We will use this information in the next section
to study the pattern of supersymmetry breaking in the exact string 
effective supergravity theory.

\newsec{Non-perturbative supersymmetry breaking in the string effective theory}
Apart from the vector multiplets, the general $\cx N=2$ supergravity
can be coupled to $\cx N=2$ matter hypermultiplets. These 
couplings may generate 
a non-trivial scalar potential which determines
the vacuum structure. We review a few facts and definitions in appendix A. 
In string theory these couplings may arise as the consequence of 
background fields such as RR fields in type II strings \Rps\ and
magnetic backgrounds in the heterotic string \Rhs. In the following
we study the contribution of instanton effects in the presence of
such backgrounds.

\subsec{The $SU(2)$ GEQFT on a compact manifold}
Let us  first consider a global type II compactification 
associated to the $SU(2)$ GEQFT as an explicit example
for the supersymmetry breaking by instantons on a compact manifold. 
It will be rather obvious how to generalize to the generic situation 
subsequently.

The prepotential of the string effective supergravity is of the form
\eqn\ppii{
\cx F(s,t)=-s(t^2+1)-a\, t^3+b\, t+c+ f(s,t),\qquad a,b,ic \in \bx R.}
The modulus $t$ is related to the single 2-sphere of the singularity
supporting the $SU(2)$ theory. In the tree level approximation 
$s\to \infty$, eq.\ppii\ is the same as \clpp\ for $G=SU(2)$ up 
to a shift of $t$ by a constant. This shift is an effect of the global 
geometry. It has no further consequences except for the translation
of the origin of the $SU(2)$ theory to $t=i$.
The function $f$ depends only on the exponentials 
$(q_s,q_t)=(e^{2\pi i s},e^{2\pi i t})$. The $s$-independent piece
of $\cx F(s,t)$ 
describes one-loop effects in the heterotic string theory
whereas a term $\sim q_s^{k}$ corresponds to an $k$ instanton effect.

The presence of background fields induces charges of a hypermultiplet 
w.r.t. to a gauge symmetry gauged by the vector multiplets.
Let us first ask what kind of gauge charges will be relevant. 
In the Super-Higgs effect, one gravitino becomes part of a massive 
super multiplet with spin content $(3/2,1,1,1/2)$ \LPW. It is natural 
to identify these two spin one vectors with the ones in the universal sector 
of the graviphoton and the dilaton $s$, $\Si=0,1$. Note that this
choice is universal and is valid for any gauge group $G$.
In addition there may be an ordinary Higgs effect in the {\it magnetic}
field theory $U(1)$ factors related to the condensation of monopoles. 

To proceed we perform a symplectic transformation 
on the standard section \standsec\ and obtain a period vector $\tx \Pi$ 
of the form 
\eqn\spvii{
\tx X^\Si=\pmatrix{
-\fc{1}{2} (t^2-1)+\fc{1}{2}\, f_s\cr
-t\cr
-\fc{1}{2} (t^2+1)+\fc{1}{2}\, f_s},\ 
\tx F_\Si=\pmatrix{
-t^2 s+a t^3-b t+s-2 c-2f+tf_t+s f_s\cr
-2 t s+3 a t^2+b+f_t\cr
t^2 s-a t^3+b t+s+2 c+2 f-t f_t-s f_s}.}
This basis is the equivalent of \clpv\ in the global compactification.
Note that the $s$-dependend piece of $\tx X^\Si$ is entirely due to the 
space-time instantons correction 
$f_s=\p f/\p s\sim q^s=e^{2\pi i s}+\dots$. Therefore in the
perturbative, tree-level plus one-loop corrected supergravity, no prepotential
exists and partial supersymmetry breaking appears to be possible. 

\subsec{Non-perturbative gravitino masses}
However, as asserted already, including the space-time instantons,
the $X^\Si$ should be always good homogeneous coordinates 
and the arguments of the no-go theorem \CGP\ apply. To be concrete,
let us consider a breaking to $\cx N=1$ near the origin $t=i$ of
the field theory Coulomb branch. The
gravitino mass matrix is proportional to the supersymmetry 
variation $S_{AB}$:
\eqn\smati{
iS_{AB}=-\fc{i}{2}e^{K/2}\, 
\pmatrix{ 
 \pp{0}{1}-i\pp{1}{1}-i\pp{0}{1}-\pp{1}{2}& 
-\pp{0}{3}+i \pp{1}{3}\cr
  -\pp{0}{3}+i \pp{1}{3}&
  -\pp{0}{1}+i \pp{1}{1}-i \pp{0}{2}-\pp{1}{2}\cr}}

\ni
A representative chocie for a breaking to $\cx N=1$ with a 
zero single eigenvalue of the matrix $S$ is
\eqn\defcharges{
\pp{0}{1}=\pp{1}{2}=m, \qquad
\pp{\Si}{x}=0\  {\rm for}\  \Si=0,1,\ {\rm else.}}
With this choice the mass matrix for generic moduli is
\eqn\smatii{
iS_{AB}=\fc{im\, e^{K/2}}{4}
\pmatrix{(t-i)^2-f_s&0\cr0&-(t+i)^2+f_s}}
For $q_s \to 0$ there is a $\cx N=1$ vacuum at $t=i$. 
For $q_s\neq 0$ the instantons lift the zero eigenvalue. Since 
the no-go theorem  applies, there must be either $\cx N=2$ or $\cx N=0$ 
supersymmetry.
An $\cx N=2$ vacuum requires $t=0,\ f_s=-1$. Even if such a point in
the moduli space would exist, it would not be connected to the large $s$ 
vacuum. Thus the instantons break further $\cx N=1 \to \cx N=0$ at a scale
$\Lambda \sim e^{-8\pi^2/g^2}$.

The exciting fact about this instanton generated supersymmetry breaking is
that the non-perturbative effects are 
determined by mirror symmetry in the geometric type II theory.
Since the non-perturbative gravitino masses arise from the coupling
of the GEQFT to the string and gravity sector, the precise instanton
series depends on the individual compactification manifold M. As an
example let us consider the mirror of the Calabi--Yau manifold\foot{
This Calabi--Yau manifold served already as a prominent example in the
understanding of typeIIA/heterotic duality \KV%
\ref\klm{A. Klemm, W. Lerche and P. Mayr, \plt 357 (1995) 313.} 
and GEQFT's 
\ref\kklmv{S. Kachru, A. Klemm, W. Lerche, P. Mayr and C. Vafa,
                \nup {459} (1996) 537.}.}
$M$ defined by the polynomial 
\eqn\cymf{
p=x_1^{12}+x_2^{12}+x_3^6+x_4^6+x_5^2
-12\,\mu\, x_1x_2x_3x_4x_5-2\,\phi\, x_1^6x_2^6.}
Here the $x_i$ are coordinates of a weighted projective space in 
which $M$ is embedded as the hypersurface $p=0$ and ($\mu,\, \phi$)
are coordinates on $\cx M_V(W)$ related to ($t,\ s$) by the mirror map.
By calculating the period integrals and using the mirror 
map\foot{We refer again to 
refs. \Rcan\ms\ for details and references 
on calculations of this type.}
we obtain the following
result for the instanton expansion $\Delta^{m(\psi)}_{np}=f_s$ 
that governs the non-perturbative
gravitino masses:

\eqn\test{\eqalign{
\Delta^{m(\psi)}_{np}\ = \  
&\qs 2\ ({\ss 2 + 2496  \qt 1+ 1941264  \qt  2 + 1327392512  \qt  3 + 
861202986072  \qt  4 + 540194037151104  \qt  5+ \dots })\ + \cr
&\qs 4\ ({\ss \fc{1}{2} + 448128  \qt  2 + 2654785024  \qt  3 + 
5718020769540  \qt  4 + 8494210810708992  \qt  5 + \dots})\ + \cr
&\qs 6\ ({\ss \fc{2}{9} +  \fc{347738368}{3}  \qt  3  
+ 2583608958216  \qt  4 + 
12741316216063488  \qt  5 +  \dots}) \  +\cr 
&\qs 8\ ({\ss \fc{1}{8} + 36401011968  \qt  4 + 2160776148604416  
\qt  5 + \dots})\ + \cr
&\dots} }
\vskip -0.3cm

\ni It would be interesting to study the global non-perturbative 
vacuum structure using this exact information. 
In fact the coefficient functions $f_k(q_t)$ of the 
$k$-th instanton term are generally modular functions
of the heterotic modular group, which is known 
for certain compactifications\foot{A study
 of the modular functions for special
cases can be found in 
\ref\Rmodg{B.H. Lian and S.-T. Yau,
Nucl. Phys. Proc. Suppl. $\us {46}$ (1996) 248;\br
M. Henningson and  G. Moore, \nup 482 (1996) 187.}.}. 
This information might be 
sufficient to determine global properties of the associated
scalar potential. In the following we will take a different
route and analyze the vacuum structure directly in an expansion
around the point particle limit of the SYM theory.

\subsec{Generalizations}
It is easy to argue that the pattern of non-perturbative 
supersymmetry breaking described in the above 
$SU(2)$ case is generic for any gauge group $G$. First note that we used 
only the universal sector of the dilaton and the graviphoton in the gauging
and the generalization to general $G$ is therefore trivial.
Consider now the electric component $\tx X^\Si$ of the
field theory section \clpv. On general grounds it is clear that the entries 
$1$ and $t_i$ will not receive $s$ dependent instanton corrections.
So the instanton corrections that lift the $\cx N=1$ vacuum are 
entirely due to the modifications of the remaining period
\eqn\ftc{
\tx X^2=\fc{\p\cx F}{\p s}=-C_{ij} t_i t_j+\dots.}
This period describes 
tree-level gauge kinetic terms, and its $s$ corrections 
non-perturbative field theory corrections to it 
plus additional non-perturbative 
gravitational and string effects in the full theory.
Thus on pure field theory grounds, the $\cx N=1$ vacuum will be lifted
in any $\cx N=2$ QFT with non-perturbative corrections coupled to 
string theory\foot{\mark We emphasize once again that the no-go theorem 
applies to the string effective supergravity, not 
to a pure field theory decoupled from string theory. \mark}.
The only
case left over are {\it conformal} QFT's with the exact gauge coupling
being equal to the tree level coupling. Still in this case we 
expect that, after adding the coupling to gravity and breaking of
conformal invariance as a consequence, 
$s$ dependent instantons appear in the
string/gravity sector\foot{It follows from the general arguments in section 2
that this is indeed the case.}. 
It is interesting to observe that the lift of the 
$\cx N=1$ vacuum seems to be closely related to the breaking of 
conformal invariance. 

\newsec{Supersymmetry breaking in the point particle limit}
In the following we want to pin down the structure of supersymmetry 
breaking as seen by the low energy observable gauge theory.
The string theory embedding predicts a special form 
of the non-perturbative 
scalar potential of the low energy gauge theory with the 
supersymmetry breaking parameters linked in a useful way 
to the string geometric moduli.

\subsec{What can be expected from the field theory}
Let us first ask what might be expected
from what is known about supersymmetry breaking in field theory. 
Obviously, since
we consider spontaneous breaking in the $\cx N=2$ supergravity theory
which encodes the exact non-perturbative $\cx N=2$ SYM theory, 
the supersymmetry breakdown should have a consistent formulation
in terms of the latter. There are three known patterns of 
supersymmetry breaking consistent with the holomorphic structure
of the $\cx N=2$ theory.

The first one is the addition of the $\cx N=1$ supersymmetric 
adjoint mass term discussed in the original work of Seiberg and
Witten \SWi. It was argued there that this term drives the theory
to the point in the Coulomb moduli where the monopole gets massless.
The combined superpotenial including the monopole hypermultiplet
$(m,\tx m)$  is 
\eqn\swsp{
W=\sqrt{2}a_D\, m\tilde m + m_{adj}\, u.}
The minimum of the scalar potential is at
$a_D=0,\ m=\tilde m=-\fc{m_{adj}}{\sqrt{2}}\ \fc{du}{da_D}$, where
the monopole becomes massless and condenses. The condensation of the 
magnetically charged monopoles leads to a mass for the 
magnetic gauge field and confinement of
the electric fields \`a la t'Hooft \Hoo.

Another very elegant and interesting 
approach to supersymmetry breaking starting from the
holomorphic properties of the $\cx N=2$ theory has been considered
in refs.%
\ref\LAG{L. \'Alvarez-Gaum\'e, J. Distler, C. Kounnas and M. Mari\~no,
\ijmp 11 (1996) 4745;\br
L. \'Alvarez-Gaum\'e and M. Mari\~no, \ijmp 12 (1997) 975;\br
L. \'Alvarez-Gaum\'e, M. Mari\~no and F. Zamora, \ijmp 13 (1998) 403;
\ijmp 13 (1998) 1847.}. Compatibility 
with the analycity properties of the exact field theory can 
be implemented by the introduction of so-called spurion fields 
with vev's that trigger the breaking. It was already argued 
there that the structure of the potential obtained in this
way is in qualitative agreement with what one expects from a 
supergravity theory.

Indeed, we will find that the supersymmetry breaking in
string theory is morally a generalization of the two 
mechanisms\foot{There are extra terms that resemble
the soft breaking terms of the $\cx N=1$ spurion 
approach 
\ref\Eva{N. Evans, S.D.H. Hsu and  M. Schwetz, \plt 355 (1995) 475;\br
N. Evans, S.D.H. Hsu, M. Schwetz and S.B. Selipsky,\nup 456 (1995)
205}.}. In particular the supersymmetry breaking parameters will depend 
in a specific way on the geometry of the transverse dimensions.

\subsec{String effective action}
To study the string effective theory near the point particle limit, 
we consider an expansion in $\al'\sim 1/\mstr^2$, 
keeping the exact quantities including the infinite instanton
series at each order. This approach has been introduced
in \kklmv\ to derive the non-perturbative field theory results
from string theory. Specifically one considers the limit of small $\eps$
with 
\eqn\ftl{
\fc{\Lambda}{\mstr}=\eps,
\qquad \Lambda\sim \mstr\, e^{-8\pi^2/b_1g^2}={\rm fixed},}
where $b_1$ is the one loop beta coefficient of the SYM theory.
Note that the tree level coupling $g$ at the string scale - equal to
the volume of the base $B$ - behaves as $\Im s \sim -b_1\, \ln \eps$.
It was further argued in \kklmv\ that in this limit the
string effective action is described by a supergravity period vector $\Pi$ 
with  building blocks
\eqn\Elowordp{
1,\ \eps\, a^k,\ \eps\, a_D^k,\ \eps^2\, u,\ s,\ \eps^2\, su,\qquad k=1,\dots,r,}
where the $a^k$ ($a_D^k$) are the vev's of the scalar fields in the
electric (magnetic) vector multiplets of the SYM theory.
A crucial fact for the following is the appearance of $u=\vp$ at order 
$\eps^2$. The dependence of $\Pi$ on $u$ 
will be generate the $\cx N=1$ supersymmetric
mass term for the adjoint scalar in \swsp.

The structure of the string effective action around the field theory 
limit turns out to be quite intricate. We will banish  details of the 
supergravity calculations in the appendix and content ourself
with an outline of the qualitative structure in the following. 
We start with the following ansatz for the symplectic section 
$\Pi=(X^\Si,F_\Si)$:
\eqn\pvg{
X^\Si=\pmatrix{f_\al+\eps^2c_\al\, u\cr \eps {\ca} a^k},\qquad
F_\Si=\pmatrix{2\eta_\al\, f_\al\,(s+\const)\cr\eps {\ca} a_D^k},}
where $\eta=-,+,+,\dots$.
The entries $X^\al,\ \al=1,\dots,2+m$ 
describe (together with the dual periods $F_\al$) 
the universal graviphoton/dilaton sector and $m$ 
extra scalar fields $t_a$ that parametrize the geometry of the 
transverse dimensions. On the other hand the entries
$X^k,\ F_k,\ k=1,\dots,r$ are associated to the $r=\rk\, G$ field 
theory periods. The expressions $f_\al,\ c_\al$ are so far
arbitrary functions of the moduli $t_{a}$ but will have to satisfy
some constraints imposed on the section $\Pi$ by $\cx N=2$ special 
geometry. The K\"ahler potential obtained from $\Pi$ has the form
\eqn\derii{
K=-ln(\cx V)+\fc{\eps^2}{\cx V}\ \kt,}
where the precise expression for $\kt$ can be found in eq.(B.1) and
the leading term is 
\eqn\deriii{\eqalign{
\cx V&=V_B\cdot \vtv,\cr
V_B&=i\,(s-\bb s+\const)\equiv \Sigma/2,\cr
\vtv&= 2\sum_\al \eta_\al |f_\al|^2\equiv 2\, \mm.}}
In fact $\cx V$ is the volume of the Calabi--Yau manifold $M$. 
Note that the field theory dependent terms are contained 
in $\kt$ and, apart from the 
inverse powers of $\mstr$, suppressed by the overall inverse 
power of the volume $\cx V$. One can distinguish two different 
geometric scales contained in $\cx V$ which will be enter the
supersymmetry breaking scales in the following. 
These are the volume of the base $V_B \sim \Sigma \sim \ln \eps$
on which the gauge fields propagate and the volume $\vtv\sim \mm$ 
of the dimensions transverse to the field theory.

\subsec{A first look at the scalar potential}

The coupling of the hypermultiplets to the vector 
multiplets\foot{See appendix A for more details.}
induces a non-trivial scalar potential $V$ for the scalars \LPW\
\eqn\spoti{\eqalign{
V=V_{\delta_ \Psi}+V_{\delta_ \xi}+V_{\delta_ \lambda}=
-3P^x_\Si P^x_{\bb \La}\, V^{\Si\bb \La}+4k^u_\Si h_{uv}k_{\bb \La}^v\, V^{\Si\bb \La}+
P^x_\Si P^x_{\bb \La}\, U^{\Si\bb\La},}}
where the three terms are due to the gravitino, hyperino and gaugino 
variations, respectively.
The couplings $P^x_\Si,\ x=1,2,3$ (and $k^u_\Si$)
describe the interactions of
the hypermultiplets $q^u$ with 
the four-dimensional vector multiplet $A^\Si_\mu$,
where $\Si$ is a gauge index and $x$ an index of the $SU(2)_R$  symmetry.
In the present context the couplings $P^x_\al$ 
in the universal dilaton/graviphoton sector parametrize
the Super-Higgs effect 
discussed in the previous section.
Since we will be interested in a region of the 
field theory moduli space where the monopoles are light, 
we must also add their couplings to the  
magnetic
$U(1)$ factors of the field theory%
\foot{We will write all following equations in the 
appropriate local magnetic 
variables which are equal to $a_D^k=\p/\p{a^k} \cx F$ in the UV region.}%
.
If $(m^i,\tx m^i),\ i=1,\dots,r$ denote 
the $r=\rk G$ monopole hypermultiplets with charges $q^i_k$,
these couplings are described by $P^x_k=Q_k^x$ with 
\eqn\kpmon{
\eps^{-2}\, Q^x_k= q_k^i\, (\bb m^i, \mt^i)\, 
\sigma^x\, (m^i,\bb \mt^i)^T.}
The matrices $U$ and $V$ depend on the scalars $z_i$ in the
vector multiplets. 
As their general form is quite involved we will restrict to
give explicit expressions along the way when needed.

\subsubsec{The cosmological term}
The leading term in the scalar potential arises from the
Super--Higgs effect in the universal sector and does not 
depend on the field theory moduli. It represents the
cosmological constant from the view of the brane world.
Its moduli dependence is described by the $\eps^0$-piece of 
the matrices $U$ and $V$:
\eqn\leadord{
e^{-K}\, V^{\al\bb \be}_0=f_\al \bb f_\be \qquad 
e^{-K}\, U_0^{\al\bb\be}=w_{\al,\rho}g^{\rho \bb \rho}\bb 
w_{\bb \be, \bb \rho}}
where we used $t_\rho=\{t_a,s\}$ to denote the non field theory moduli
and $w_{\al,\rho}=K_\rho f_\al +f_{\al,\rho}$.
To avoid a cosmological constant of the
order of the string scale, the sum of the leading contributions 
from the gravitino, hyperino and gaugino variations in \spoti\ should vanish.
There are various possibilities to achieve such a cancellation
at special values of the vector and/or hypermultiplets.

In the following we will mostly separate the question 
of why the cosmological constant is so small compared to $\msusy$ 
from the study of the pattern supersymmetry breaking on the brane
as a function of the couplings $P^x_\al$ that determine 
the Super--Higgs effect.
Note that providing a solution to the cosmological constant 
problem in the present context would be rather significant as 
even non-perturbative effects are included. Although we will
not solve the problem of the cosmological constant 
we will observe some interesting interrelations between 
the supersymmetry 
breaking on the brane and the cosmological term and keep an eye on
possible mechanisms for a cancellation of the latter.

\subsubsec{The $\cx N=1$ preserving adjoint mass term}
Let us consider for a moment a naive cancellation of the individual
contributions to the cosmological term by imposing
\eqn\simpcond{
P_\al^x \cdot w_{\al,\rho}=0,\qquad P^x_\al\cdot f_\al=0.}
Assuming that eq.\simpcond\ holds, one finds for the 
leading piece of the scalar potential in the field theory limit:
\eqn\spneofin{\eqalign{
\eps^{-4}\, V=&\ \fc{1}{2}\, \sum_x\, 
(Q^x_l+m^x_{adj}\, u_l)\, \tau^{l \bb k} \, 
(Q^x_k+\bb m^x_{adj}\, \bb u_{\bb k})\cr
&\ + 2\,  q^i_lq^i_k\, a^l a^{\bb k} (|m^i|^2+|\mt^i|^2),}}
where 
\eqn\madj{
m^x_{adj}=M_{str}\cdot \fc{P^x_\al c_\al}{\ca} .}
Eq. \spneofin\ describes precisely the scalar potential 
of the $\cx N=2$ field theory with a
mass term $\sim m_{adj}$ for the adjoint scalars in the vector multiplets. 
As argued in ref.\SWi, the
$\cx N=2$ theory is driven to the point $a^l=0,\ \forall l$ 
in the Coulomb moduli, where $r$ mutually local monopoles
become massless and condense with a vev $\sim \sqrt{m_{adj}\Lambda}$ 
in a new $\cx N=1$ supersymmetric vacuum. 

\subsubsec{Relations from special geometry and breaking to $\cx N=0$}
Imposing the condition \simpcond\ not only cancels the leading 
order contribution to the scalar potential but also a number 
of extra terms at $\cx O(\eps^4)$. As the expression 
in eq.\spneofin\ is the most general one compatible with 
$\cx N=1$ supersymmetry these terms must necessarily break 
$\cx N=1\to \cx N=0$ in the SYM theory. 

An generic solution to \simpcond\ is the limit
of infinitely large extra dimensions, as will be discussed below.
It might appear that partial supersymmetry breaking to 
$\cx N=1$ with the adjoint mass term \madj\ is possible 
for any such  solution.
However this is in fact not true. The 
reason is that the general ansatz \pvg\ for the supergravity 
section $\Pi$ must be  subject to further conditions 
in order that $\Pi$ is a valid symplectic section of $\cx N=2$
special geometry. In the appendix B we find that these relations
have the important implication
\eqn\noneo{
P^x_\al\cdot  f_\al = 0 \ \ \Rightarrow \ \ m^x_{adj}\sim P^x_\al\cdot  c_\al = 0.}
\ni
Thus cancellation of the $\cx N=1$ supersymmetry breaking terms 
at $\cx O(\eps^4)$ implies also the cancellation of the leading
$\cx N=1$ adjoint mass term. Therefore,
contrary to what might have been naively expected,
there will be no excessively large separation of the two breaking 
scales to $\cx N=1$ and $\cx N=0$, respectively. 

\mark
As we pointed out already, the no-go theorem in the string
effective supergravity does not imply a no-go theorem for
the pure field theory completely decoupled from string theory.
This decoupling can be achieved by taking a strictly infinite
string scale $\mstr=\infty$ which, by eq.\ftl, amounts to 
a non-compact limit in the (tangential) direction of the {\it base}.
In this case certain terms dressed by inverse powers of
$\Sigma \sim \ln \Lambda/\mstr$ drop from the effective 
field theory potential and 
eq.\simpcond\ can be replaced by a weaker condition that can be
satisfied and connects to the $\cx N=1$ pure field theory 
vacuum discussed in \TV. However note that in the 
field theory coupled to string theory
the phenomenologically acceptable regime is 
$\ln \Lambda/\mstr\, {\sim\atop \ }
\!\!\!\! {\scriptstyle <}\ 20$ and supersymmetry {\it is }
broken by the terms with extra powers of 
$\Sigma$ (and in fact we will argue that the 
phenomenologically interesting region is in the range of 
moderately small $\Sigma$). 
This leads again to 
a quite restricted hierarchy for the breaking scales
to $\cx N=1$ and $\cx N=0$, respectively 
\mark

In appendix B we collect a few additional relations derived
from special geometry which contain some useful information.
In particular they imply that the scale of the supersymmetry breaking
is given by 
\eqn\condscale{
m_Q \sim b_1 \cdot \fc{P_\al f_\al}{\mm},}
which is  also equal to the  scale of the 
monopole condensate. The fact that $m_Q$ is 
proportional to the one-loop beta function coefficient 
fits well the discussion in section 4.3 where we observed
that the supersymmetry breakdown is apparently linked 
to the violation of conformal invariance. Moreover 
eq.\condscale\ displays the important fact that the 
supersymmetry breaking scale tends to be suppressed
by the volume of the transverse dimensions. 
\goodbreak

\subsubsec{Non-compact transverse dimensions}
The defining data of the string effective
theory is is the period vector $\Pi$ of a K3 fibered 
Calabi--Yau manifold. A special property of the 
leading periods $f_\al$ in the universal 
sector is that they are free of any instanton corrections 
arising from a finite transverse volume $\mm$. The reason is 
that these $s$-independent pieces are identical to the periods
of a K3 manifold which is of the same type as the generic fiber
in the fibration. Moreover on the K3 manifold there are no 
instanton corrections to the period integrals as a consequence of
the $(4,4)$ supersymmetry of the CFT. Explicitly this
means that the periods $f_\al$ are polynomial in the $t_a$ of
degree two with the precise form dictated by the K3 intersection
matrix. Note that there {\it are} instanton corrections 
with an action proportional to the K\"ahler volumes $\{t_a\}$ 
in the K3 {\it fibration}. 

Obviously, a solution to both conditions in \simpcond\ 
can not exist for generic moduli. It would be interesting to
classify possible solutions on sub-slices of the moduli of all K3's,
which would be tantamount for finding solutions for a
vacuum with vanishing leading cosmological constant. 
In the next section we will study in detail 
the dependence of the scalar potential on the 
universal volume modulus $t$ of the transverse dimensions.
In this case it is easy to verify that the only solution to \simpcond\ 
is the limit of infinitely large
transverse dimensions, $t=\infty$. For the reasons 
described above this does not
lead to partial supersymmetry breaking, however.
\mark
Note that non-compact transverse volume should not be confused with 
the pure field theory case, which requires infinitely large 
tangential dimensions $V_B\sim \ln \Lambda/\mstr \to \infty$.
\mark

\newsec{Supersymmetry breaking on the brane}
After having outlined some general features we give now
details of the supersymmetry breaking 
in the effective SYM theory on the brane as a function of 
the couplings $P^x_\al$ parametrizing the Super--Higgs effect
in the universal string sector\foot{Note that the following
expressions describe the supersymmetry breaking on the brane
for arbitrary general couplings and thus the distinction of patterns
with partial or complete classical Super-Higgs effect in the string sector
arises only after making a specific choice for them.}. For concreteness 
we will restrict the description
of the deformations of the transverse geometry to the generic 
volume modulus $t$. Introducing extra moduli will not change 
the effective theory in the sector of the SYM theory;
the way that these extra moduli enter is in the leading, cosmological
term.

The universal sector of the graviphoton, the dilaton
and the universal modulus $t$ is described by the following 
ansatz for the functions $f_\al$:
\eqn\falfs{
f_0=\fc{1}{2}\, (1+t^2),\quad 
f_1=\fc{1}{2}\, (1-t^2),\quad 
f_2=t.}
Moreover it follows from the relations of special geometry that 
the coefficients of the $u$ dependent terms in the 
symplectic section \pvg\ are
\eqn\cis{
c_0=-b,\quad c_1=b,\quad  c_2=0,\quad \bb c_0=-c_0,}
where $b\sim b_1$ with a numerical coefficient specified in (B.5).
With this information one can determine the matrices $U$ and $V$ 
and the scalar potential eq.\spoti. Since its general form 
is at first sight involved, we will perform the analysis 
in various steps.

\mark
\subsec{Leading terms in a large $\Sigma$ expansion}
Let us start with the leading terms of the
field theory potential in an expansion in powers of 
\eqn\eqsig{
\Sigma=-\fc{b_1}{\pi}\ln \Lambda/\mstr.}
The motivation for this expansion is that for a very large string scale,
the terms with lower powers of $\Sigma$ are suppressed\foot{In 
fact the relative contribution of these terms might be
underestimated by the naive formula \eqsig, since it does 
not take into account threshold effects which 
are known to be sizable in string theory compactifications
with large compact dimensions.}. In particular
the pure field theory potential will correspond to keeping
only the leading terms in the limit $\Sigma=\infty$.
However for the theory coupled to gravity, the logarithmic 
dependence on $\mstr$ implies relative small values of $\Sigma$
bounded from above by the value $\sim 10\ b_1$ for $\mstr \sim M_{pl}$.
In fact, as we will argue that the supersymmetry breaking scale
is proportional to the string scale and moreover the latter
can well be in the TeV range in this ``world on a brane'' scenario,
the value of $\Sigma$ can be actually quite small for a
moderate to low value of the string scale. In this case
the terms with lower powers in $\Sigma$ are relevant.
However we find it still useful to display the
structure of the potential by organizing it in powers of $\Sigma$,
keeping in mind that 
this will only give a  good physical hierarchy for extremely
large values of $\mstr$.
\mark

From the general expressions in appendix C one obtains the
following  leading 
term of the scalar potential in the SYM theory:
\def\mz{m_{_{\cx N=0}}}\def\mzb{\bb m_{_{\cx N=0}}}
\def\mo{m_{_{\cx N=1}}}

\eqn\genspoti{\eqalign{
\eps^{-4}\, V=&\ \
\fc{1}{2}\{Q^-_l+M^-_l\}
\, \tau^{l\bb k} \,
\{Q^+_k+M^+_k\}+
\fc{1}{2}\{Q^3_l+M^3_l\}
\, \tau^{l\bb k} \,
\{Q^3_k+M^3_k\} \eqsp\cr
&+\, 2\,  q^i_lq^i_k\, a^l a^{\bb k} (|m^i|^2+|\mt^i|^2) \eqsp\cr&
-\, \alpha\fc{ib}{2\mm^2}\,   (u_s-\bb u_{\bb s})
+(\fc{i}{2\mm} \beta u_s + \cc). \eqsp
}}
Here the superscripts $\pm$ refer to complex combinations $x_1\pm ix_2$
of the real fields. In particular $Q^-_l=2m^i\mt^iq^i_l$ is the 
holomorphic bilinear in the monopole fields and $Q^+_l=(Q^-_l)^*$.
The supersymmetry breaking terms in the potential can be
divided into 
\vskip 0.2cm

\leftskip=15pt
\ni
$\underline{i)\ F-type\ terms}$:\br
The vev of the holomorphic bilinear term in the monopole fields 
will be determined by the quantity $M^-_l$ which can be
split into its holomorphic and harmonic pieces in the SYM fields,
respectively:
\eqn\massterms{
M^-_l=\mo u_l + \mz (u_l-\cc).}
The holomorphic piece 
\eqn\mneo{
\mo =\fc{b}{\bb \ca\mm}P^-_\al(f_\al-\bb f_\bal),}
represents the $\cx N=1$ preserving adjoint mass term
depending on the transverse volume as dictated by \falfs.
On the other hand, the harmonic piece
\eqn\mnez{
\mz=\fc{b}{\ca\mm}P^-_\al f_\al}
induces a soft breaking of supersymmetry.

\vskip 0.2cm
\ni
$\underline{ii)\ D-type\ terms}$:\br
The alignment of the vev's of the scalar components of the 
individual chiral $\cx N=1$ multiplets in the 
monopole hypermultiplet are 
determined by the remaining component $M^3_l=\zeta u_l+\cc$ with 
\eqn\dterm{\zeta=\fc{b}{\ca\mm}P^3_\al\bb f_\bal.}
\vskip 0.3cm

\vskip 0.2cm
\ni
$\underline{iii)\ Remaining\ terms}$:\br
In addition there are further soft-breaking terms that depend on 
$u_s$ with coefficients
\eqn\sbterms{\eqalign{
\alpha=\fc{\mm^2}{b^2}|\ca|^2(\fc{1}{2}|\mz|^2+\fc{1}{2}|\mz+\mo|^2+|\zeta|^2),\cr
\beta=\fc{\mm}{b}(\fc{1}{2}\ca(\mo+\mz)\, C^++\fc{1}{2}\bb \ca \mzb C^--\ca\zeta C^3),}
}
where $C^x=P^x_\al c_\al$. A nice heuristic interpretation of
these terms can be given as follows.
Let us consider the $\cx N=1$ supersymmetric term in
\massterms. In the field theory this term can be associated with
a superpotential term $u\, \mo$ in eq.\swsp. 
At the point where the $\cx N=1$ vacuum branches 
off, $u=\omega_G\Lambda^2$,
where $\omega_G$ is a $c_2(G)$-th root of unity. Moreover 
the scale of the $\cx N=1$ theory is related to that of the $\cx N=2$
theory by 
\def\Lo{\Lambda_{_{\cx N=1}}}
$\Lo^3=\mo\Lambda^2$. Thus the superpotential term $u\, \mo $,
evaluated at the $\cx N=1$ point of  the Coulomb branch is
\eqn\ahw{
W=\omega_G\Lo^3.}
So remarkably enough, evaluating the superpotential
of ref.\SWi\ at the extremum with respect
to the SYM fields, gives precisely the expected dynamical
generated superpotential in the $\cx N=1$ theory. In the string theory
context there are further contributions from the superpotential \ahw\ to the
scalar potential because $\Lambda\sim e^{i\gamma s}$ 
is to be treated as a field rather than a constant. These
terms have the form of the $u_s$ dependent terms in the scalar potential
\genspoti.

\vskip 0.3cm

\leftskip=8pt
\ni
To determine the scale of supersymmetry breaking, note that
$\p \tau^{l\bb k}/\p a^n$ is very large in a region with 
light monopoles, $a^k\sim 0$. As a consequence, minimization with respect 
to the fields $a^k$ will require an adjustment of the monopole
bilinears $Q^x_l\sim M^x_l$ to a high precision. Thus
the generic scale of the monopole condensate and the 
supersymmetry breakdown is given by 
\eqn\sbsc{
\msusy\sim \fc{b_1}{\mm}\, P^x_\al f_\al \mstr \sim \fc{b_1}{\mm}\mstr,}
where in the second expression we have assumed that the couplings
in the universal string sector are of order one in string units. 
As we will discuss in the next section, this is indeed the case 
if these couplings $P^x_\al$ arise  from certain background fields in the
string theory compactification.

\subsubsec{The $\cx N=1$ pure field theory vacuum\foot{I thank
Cumrun Vafa for conversations on the following issues.}}
If we are not interested in string theory but pure field theory,
we can take the decoupling limit of infinite string mass 
which implies $\Sigma\to \infty$ by eq.\ftl. Only the 
leading terms \genspoti\ survive in this pure field theory case.
This case was considered in \TV\ and it was asserted that 
a solution with $\cx N=1$ supersymmetry exists. Note that there
are many choices for the Super--Higgs effect that lead to 
partial supersymmetry breaking in the classical supergravity
and break the remaining supersymmetry only by instanton effects.
However only a very special subset  
will be related to a vacuum of the $\cx N=1$ pure field theory
at $\mstr=\infty$.

A choice of or RR-fluxes (or Super--Higgs effect) 
that cancels all supersymmetry
breaking terms in the leading part of the potential must correspond to
$\zeta=\mz=0$ and values for $\alpha$ and $\beta$ such that the
last terms in \genspoti\ cancel. From eqs.\mnez\ -- \sbterms\ we find
that the solution is given by
\eqn\neofl{
P^-_0=-\fc{1+t_0^2}{1-t_0^2}\ P^-_1,\qquad
P^-_2=\fc{2t_0}{1-t_0^2}\ P^-_1,\qquad
P^3_\al=0, \  \forall \al.}
The meaning of this equations is that for fixed
$t_0$, the choice \neofl\ of the 
Super--Higgs cancels the supersymmetry
breaking terms in the leading potential at $t=t_0$.

Let us summarize some relevant properties of the solution \neofl\ that 
represents a special configuration of 
RR-fluxes that preserve $\cx N=1$ in the decoupled field theory 
at $\mstr=\infty$. 
({\it 1}) It corresponds to a special 
fine tuning of fluxes dual to 0, 2 and 4-form charges 
on the non-compact manifold. 
The leading contribution in the non-compact transverse 
limit relevant for geometric
engineering of $\cx N=1$ QFT's is from the flux dual to the 0 and 4-form
charges
(this outcome is slightly different from the 2-form flux proposed in \TV).
({\it 2}) The solution exists already at finite $t$.
The relevant
decoupling is due to the infinite volume of the 
dimension {\it parallel} to the brane, $\Im s\sim \ln \Lambda/\mstr\to \infty$
which gives an infinite mass to the string states. 
This is in nice agreement with our previous assertion
that the breaking to $\cx N=0$ is indeed only due to the $s$-dependent
instantons in the string sector, which are non-perturbative in the 
field theory coupling (rather than instantons from the, in string units, 
finite transverse volume). ({\it 3}) For finite $\mstr$ there are 
extra terms discussed in the next section that come with additional
powers of $1/\Sigma$ and break
supersymmetry. Canceling the leading terms by \neofl\ leads to an
extra logarithmic dependence of the supersymmetry breaking scale on
$\mstr$. It is interesting to mention that the overall coefficient
of these subleading terms depends on the hypermultiplet involved
in the Super--Higgs effect (the value of $\lambda$ defined in the
next paragraph). For the hypermultiplet involved in the string
theory Super--Higgs effect this coefficient is non-zero 
\Rps\ref\Mich{J. Michelson, \nup 495 (1997) 127.}. If the 
coefficient would have been zero, supersymmetry breaking would
have been postponed to the next order of $\Lambda/\mstr$. 
\mark

\subsec{Subleading terms and a vanishing leading cosmological constant}
Subleading terms in the $\Sigma$ expansion can be straightforwardly added
using the expressions in appendix C. For the purpose of 
displaying their structure however it is useful to temporarily  make 
a concrete and simple choice for the Super--Higgs effect. 
For a  physically motivated illustration let us cancel the leading 
cosmological constant in the large volume limit by imposing 
eq.\simpcond\ on the leading terms. This is achieved by 
\eqn\chargechoice{
P^1_0=P^1_1\equiv m/\mstr,\qquad P^x_\al=0, \ {\rm else},}
were we have used a $SU(2)$ rotation to fix the direction of the 
triplet of Killing prepotentials. To treat the hyperino contribution
in \spoti\ on the same footing with the gaugino and gravitino we will
use the relation  $k^u_\Si h_{uv} k^v_\La= \lambda P^x_\La P^x_\Si$.
This simplification is justified for the couplings in string 
theory with $\lambda=4$, as will be outlined in the next section. 
The full field theory dependent scalar potential for the 
Super-Higgs determined by \chargechoice\ is 

\eqn\fullsp{\eqalign{
\eps^{-4}\, V=
&\ \fc{1}{2}
\{Q^1_k-\fc{\hx M}{|{\ca}|^2}\,u^I_k\}
\ \tau^{k\bb l}\ 
\{Q^1_l-\fc{\hx M}{|{\ca}|^2}\, u^R_l\}
+\fc{1}{2} \tau^{k\bb l}\, (Q^2_kQ^2_l+Q^3_kQ^3_l)
\eqsp\cr&
\ +\{\gamma\, (2u-a^lu_l)\fc{i \hx M^2}{2b}+\cc\}+
2\, (a^lq_l^i)\, (\bb a^{\bb k}q_k^i)\, (|m^i|^2+|\tx m^i|^2)
\eqsp\cr&
\ -\{\fc{2M^2}{b\Sigma}(1-\fc{2i}{\gamma\Sigma})\, u+\cc\}
\eqsp\cr&-
\ \{2 {\ca}\fc{M}{b\Sigma}\, Q^1_la^l+\cc\}
+2i|{\ca}|^2\fc{M^2}{b^2\Sigma^2}(a_D^l\bb a^l-\cc)
+\lambda\fc{M^2}{b^2\Sigma^2}\kt
.\eqsp}}

\ni
Here $u^I_k={\ca}\, u_k+\bb {\ca}\,\bb u_{\bb k}$ and 
\eqn\MMdef{
M=\fc{b}{\mm}\fc{m}{\Lambda},\qquad \hx M=(1-\fc{4i}{\gamma\Sigma})\, M.}
Let us compare this result with the soft breaking terms discussed 
in the context of SYM theories in the literature.
The first two lines of eq.\fullsp\ 
have the form of soft breaking terms generated by the spurion
approach of \LAG. The term in the third line is of the form generated
by the $\cx N=1$ spurion of \Eva. The coefficients of these
breaking terms are related in the string effective theory 
which can be thought of explicitly determining the
vev's of the spurion fields. The fourth line contains
some additional soft breaking terms. 

Note that the $\cx N=1$ adjoint mass term is absent only 
because of 
the choice of couplings \chargechoice\ and as a consequence 
of eqs.\madj\ and \cis.

\subsec{Minima of the potential}
Ultimately, all moduli including 
the numerical values of $\Sigma$ and $\mm$ 
should be fixed by the minima of the potential. 
The minimization with respect to the field theory moduli
is relatively straightforward and reduces  approximately
to $Q^x_l\sim M^x_l$ with small corrections from the 
additional finite terms near the monopole point.
In the special case with $\mo=0$ above, the leading potential in 
a large $\Sigma$ expansion is of the form of 
the soft supersymmetry breaking considered in refs.\LAG,
which include a detailed numeric study of the  vacuum structure
for the gauge group $SU(2)$. 
By comparison, the spurion vev should 
be identified with the scale $\msusy$ in \sbsc.

To determine the string vacuum, one needs also to vary with respect to 
the dilaton $s$ and the transverse volume modulus $t$. 
This brings us back to the question of the cosmological 
term. We will not solve the question of minimization for these 
fields, which we leave for the future,
but point out some qualitative features. As for the dilaton,
a stabilization is possible due to the subleading terms in $\Sigma$
with explicit $s$ dependence. Note that there are also further
calculable terms at order $\eps^6$ and higher which depend non-trivially
on $s$. Thus although we do not know in the moment which value of $s$
corresponds to a minimum, there is no obstruction in principle
to answer this question. 

A similar situation holds for the volume modulus $t$ of the large
extra dimensions. For a fixed $\msusy\sim$ 1TeV, the behavior of 
the typical inverse radius vs. $\mstr$ is plotted in the figure below.

{\baselineskip=12pt \sl
\goodbreak
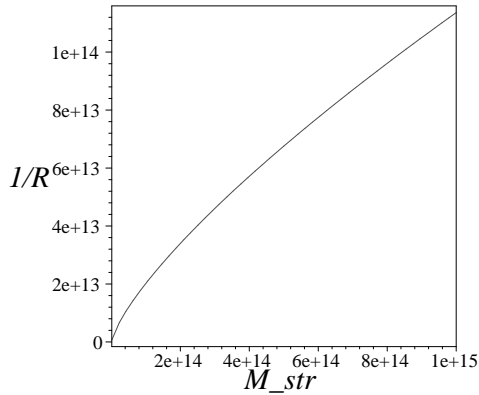
\midinsert
\centerline{\epsfxsize 2truein \epsfbox{fig1.eps}}
\leftskip 1pc\rightskip 1pc \vskip0.3cm
\noindent{\ninepoint  \baselineskip=8pt 
}\endinsert}
\vskip-2cm

\ni In particular, for small a $\mstr$ in the TeV range, the 
average radius is close to $\mstr^{-1}$ while it is larger 
by a factor of $\sim 10^{4}$ 
for a string scale equal to the Planck scale.
Since the scale of supersymmetry breaking is 
$\msusy\sim \vtv^{-1}$
one suspects there is no local minimum for the volume modulus and 
thus $t$ will run away to infinity to restore 
supersymmetry. A natural idea to fix the minimization of 
the transverse volume is 
to link it to the minimization of the field theory moduli space
\ref\Rip{Work in progress.}.
In other words, we can use, as in the compactification discussed 
in section 4, 
the fact that the volume modulus itself can correspond to one of the
vev's of the field theory gauge group. Since the transverse volume
is of order one in string units in such a model, one expects a 
relation 
\eqn\msmsi{\msusy \sim \mstr \sim R^{-1}.}
Therefore this class of compactifications requires the string scale
to be equal to the supersymmetry breaking scale in the TeV range
and predicts it to be equal to the scale of the extra large dimensions.

\newsec{Geometric charges in string theory and why $\mstr$ should be small}

In the previous sections we have described the supersymmetry
breakdown as a consequence of a Super-Higgs effect 
in the universal sector, 
parametrized by the couplings $P^x_\al$. Let us finally
comment on some important details of the relation of these couplings 
to the background fields in string theory.

In ref.\Rps\ it was shown that backgrounds of RR 2$p$-forms in the 
type IIA theory induce charges of the dilaton hypermultiplet 
w.r.t. a gauge symmetry of the vector multiplets specified
by the cohomology class of the background. From the 
equations in \Rps\ it also follows that $\lambda=4$ for this case.
Due to the special property of the RR sector in the type II 
theory these couplings are suppressed by a factor of the 
string coupling $\lambda_{II}$.

This is an extremely important detail from the world on a brane 
point of view since $i)$ the string scale $\mstr$ may be 
in the TeV range in this scenario\ref\Rap{I. 
Antoniadis and B. Pioline, \nup 550 (1999) 41.}; 
$ii)$ the supersymmetry breaking
scale \sbsc\ is of the order of $\mstr$ rather than $M_{pl}$.
In fact the relation between these two scales is given by dimensional
reduction as 
\eqn\plscales{
M_{pl}^2\sim\fc{\Sigma \mm}{\lambda_{II}^2}\mstr^2.}
The special origin of the SYM theory - its origin in the RR sector
which makes its coupling independent of $\lambda_{II}$ and
its localization on a hypersurface which makes its coupling 
independent of the transverse volume $\mm$ - has the consequence
that the values of $\mm$ and $\lambda_{II}$ remain largely undetermined
by phenomenological requests. This allows for a large separation
of the string scale and the Planck scale
arising from a large transverse volume $\mm$ and/or 
a small string coupling $\lambda_{II}$. 

After supersymmetry breaking, even a small 
mismatch in the cancellation of the cosmological constant 
in the field theory will drive the string scale to as 
small a value as possible. In other words not only is it
consistent to have strings at the TeV scale in this
world on a brane scenario, but, after the supersymmetry breakdown, 
a low string scale may well be energetically favored.
In fact the solution to the 
cosmological constant problem should tell us 
why $\mstr \sim$ a few TeV instead of being zero.

\subsubsec{Hidden sectors}
A type IIB version of the backgrounds of ref.\Rps,
with a generalization to also include NS backgrounds
has been given in refs.\Mich\TV\foot{To compare
the results of these two papers, which in fact seem to disagree in 
that the scalar potential of the \TV\ is $SL(2,\bx Z)$ invariant
whereas the one of \Mich\ is not,
one needs to rewrite the 
$\cx N=2$ scalar potential in terms of a $\cx N=1$ superpotential,
a task that appears to have been settled only in very special cases, see 
\ref\Rneosps{J.P. Derendinger, S. Ferrara, A. Masiero and A. Van Proeyen, 
\plt 140 (1984) 307;\br
E. Cremmer, C. Kounnas, A. Van Proeyen, J.P. Derendinger, S. Ferrara, 
B. de Wit and  L. Girardello, \nup 250 (1985) 385.}. We 
propose the  general form of the $\cx N=1$ superpotential 
to be $W=P_\Si X^\Si- \tx P^\Si F_\Si$, where 
$P_\Si=P^1_\Si+iP^2_\Si$ is the complex combination of the real
Killing prepotentials and we have included a similar term $\tx P^\Si$ 
for the couplings to the magnetic fields to ensure
symplectic covariance. To verify the consistency of this ansatz,
the use of the
supersymmetric Ward-identity in eq.(2.97) of ref.%
\ref\DFF{R. D'Auria, S. Ferrara and P. Fre, \nup 359 (1991) 705.}
is essential.}.
In ref.\TV\ the coupling of the universal hypermultiplet to the vector
gauge symmetries was derived from the BPS tension of 
NS and RR 5 branes in the presence of such backgrounds. 
As the tension of the NS brane lacks
the suppression factor $\lambda_{II}$, NS backgrounds will naturally
lead to a supersymmetry breaking scale $\sim M_{pl}$. 
A low supersymmetry breaking scale could still be obtained
by breaking first supersymmetry on a ``hidden singularity'' and
then transmitting the breakdown by gravitational interactions
to the observable branes, similarly as in refs.\gc.

\subsubsec{Outlook}
Using the appropriate geometric singularities of ref.\KMV, 
there are no obvious obstacles to model a $\cx N=2$ supersymmetric 
string theory embedding of the gauge group of 
the standard model on a brane and 
its matter spectrum. Breaking supersymmetry in the way
described in this paper leads to a generalized, calculable 
scalar potential depending on the parameters of this 
``standard model''. Apart from 
questions of details, e.g. how natural the standard model
spectrum appears in the context of geometric singularities,
higher derivative corrections etc, 
this appears to be an unexpectedly 
accomplishable scenario for the study of a phenomenologically
relevant string vacuum
with calculable, dynamically fixed mass and 
coupling parameters. 

\vskip 2cm
\ni
{\bf Acknowledgments}:\br
I am grateful to  Sheldon Katz, Wolfgang Lerche and Cumrun Vafa 
for essential comments and conversations.
I would also 
like to thank
Luis \'Alvarez-Gaum\'e,
Costas Kounnas,
Hans-Peter Nilles,
Antoine van Proeyen
and Stephan Stieberger
for valuable discussions and comments.

\appendix{A}{A brief review of the $\cx N=2$ scalar potential}
Apart from the vector multiplets, the
$\cx N=2$ supergravity may contain a number $n_H$ of matter
hypermultiplets $q^u$,
with the $4\ n_H$ scalars components parametrizing a 
quaternionic manifold $\cx M_H$. The $\cx N=2$ supergravity allows 
for couplings of the hypers to the vectors by gauging 
isometries of $\cx M_H$.
In other words, hypermultiplets can be
charged under the $U(1)^{n_V+1}$ gauge symmetry
\eqn\hmgauging{
q^u \to q^u + k^u_\Si\eps^\Si,}
with the Killing vector $k^u_\Si$ defining the charge of $q^u$
under the $\Si$-th $U(1)$ symmetry\foot{In particular FI terms can be 
included in this description by gauging a ``trivial'' hypermultiplet.}.
We refer to 
\ref\ABCDFF{L. Andrianopoli, M. Bertolini, A. Ceresole, R. D'Auria, 
S. Ferrara and P. Fr\'e, J. Geom Phys. $\us {23}$ (1997) 111;
\nup 476 (1996) 397.}
for a most general and modern account of the 
combined $\cx N=2$ effective action and a detailed list of references. 

The supersymmetry variations 
of the gauginos $\lambda^{\bb a}_A$, the hyperinos $\xi^\al$ and the
gravitino $\psi_{A\mu}$ depend on the scalars $z_i$ as:
\eqn\susyvi{
\delta \lambda^{\bb a}_A=W_{AB}^{\bb a} \ \eta^B,\qquad
\delta \xi^\al=N^\al_A \ \eta^A,\qquad
\delta \psi_{A\mu}=iS_{AB}\gamma_\mu \eta^B,}
with
\eqn\susyvii{\eqalign{
W_{AB}^{\bb a}&=ie^{K/2}(\sigma^x\eps)_{AB}P^x_\Si(\p_b+\p_b\, K)
X^\Si g^{b\bb a},\cr
N^\al_A&=2U^{\al B}_u\eps_{BA}e^{K/2}k^u_\Si X^\Si,\cr
S_{AB}&=-\fc{1}{2}(\sigma^x\eps)_{AB}P^x_\Si e^{K/2}X^\Si.}}
Here $A=1,2$, $\Si=0,\dots,n_V$, $a=1,\dots,n_V$,
$K$ is the K\"ahler potential, $U^{\al B}_u$ the symplectic vielbein 
and moreover the $P^x_\Si,\ x=1,2,3$ a triplet of Killing prepotentials 
associated to the 
gauging \hmgauging. 

The coupling \hmgauging\ of the hypermultiplets to the vector multiplets
induces a non-trivial scalar potential $V$ for the scalars \LPW\
which can be written in terms of the supersymmetry variations as 
\eqn\aspoti{\eqalign{
V=&-6\, \tr\, S\, S^*+
\tr\, N^{\dagger}\, N
+\fc{1}{2}g_{a\bb b}\, \tr\, (W^{\bb a})^ *\, W^{\bb a}=\cr
&(-3P^x_\Si P^x_{\bb \La}+4k^u_\Si h_{uv}k_{\bb \La}^v)\, V^{\Si\bb \La}+
P^x_\Si P^x_{\bb \La}\, U^{\Si\bb\La},}}
with 
\eqn\aspotdefs{
V^{\Si\bb\La}=e^K\, X^\Si \bb X^{\bb \La},\qquad 
U^{\Si\bb\La}=f_i^\Si g^{i\bb j} \bb f_{\bb j}^{\bb \La},\qquad
f^\Si_i=e^{K/2}\, (\p_i+K_i)\, X^\Si.}

\appendix{B}{Special geometry of the string effective GEQFT's}

To determine the effective string theory we need the precise form 
of the section $\Pi$. The large $s$ behavior \clpp\ and the 
consistency of the supergravity monodromies with the monodromies
of the embedded field theory imply the general form \pvg\
$$
X^\Si=\pmatrix{f_\al+\eps^2c_\al\, u\cr \eps {\ca} a^k},\qquad
F_\Si=\pmatrix{2\, \eta_\al f_\al\,(s+\const)\cr\eps {\ca} a_D^k}.
$$
The K\"ahler potential has the $\eps$ expansion
$$
K=-\ln\ i(\bb X^\Si F_\Si- X^\Si \bb F_\Si)=-ln(\cx V)+\fc{\eps^2}{\cx V}\ \kt,
$$
with $\cx V$ given in \deriii, $A=\sum \eta_\al \bb f_\al c_\al$ and
\eqn\defkt{
\kt=i|{\ca}|^2(a^k\bb a_D^k - \bb a^k a_D^k)
+2i(A\bb s u -\cc)+h(u,s)+\bb h(\bb u,\bb s).}
The moduli dependent quantity
$\mm=\sum_\al \eta_\al\, | f_\al|^2$
describes the volume of the transverse dimensions.
From the above expressions we can determine the matrices 
$U$ and $V$ entering the scalar potential \spoti.

\subsubsec{Special geometry of  $\Pi$}
Special geometry implies some important conditions 
on the functions appearing in the ansatz for $\Pi$ \pvg.
From the supergravity identity
\eqn\sht{
U^{\La\bb \Si}=-\fc{1}{2}(\Im \cx N)^{\La\Si}+V^{\La\bb \Si},}
and the fact that $U$ is hermitian we see that 
$U-V$ must be a real symmetric matrix.
Let us consider the leading $\eps^1$ term of $U^{\bb \al k}$. After
some algebraic manipulations we obtain

\eqn\uti{\eqalign{
U^{\bb \al k}_1&=V_1^{\bb \al k}
+A^\al g^{\bb kl} u_l
+B^\al g^{k\bb l}  \bb u_{\bb l},\cr
A^\al=
&-\fc{{\ca} b}{\cx V \mm}\{
\cx V \bb f_\bal + 
\fc{2i}{\gamma}\bb w_{\bal, \bb a} g^{\bb a b}\mm_b
\},\cr
B^\al=
&-\fc{{\ca} \bb A}{\cx V \mm}\{\cx V (\bb f_\bal-\fc{\mm}{\bb A}\bb c_\bal) - 
\bb w_{\bal, \bb a} g^{\bb a b}\, (
2is\mm(\fc{\bb A_{b}}{\bb A} -\fc{\mm_b}{\mm})+\cr
&\qquad \qquad \ \ 
\fc{\mm}{\bb A}\, (\fc{\mm_b}{\mm}\bb h_{\bb u}-\bb h_{\bb u b})
\},
}}
where $w_{\al,a}=K_a\, f_\al+f_{\al,a}$. Moreover $b$ and $\gamma$ are 
two constants that are defined in the field theory:
\eqn\ftppii{\eqalign{
\cx \p_s F_{FT}=&-\fc {2}{|{\ca}|^2}\, b\, u,\qquad \Lambda=e^{\gamma s},\cr
\cx F_{FT}=&\fc{1}{2}\sum_{\rk  G} a^ka_D^k+\fc{ b}{|{\ca}|^2\gamma}\, u.
}}
In particular the constant $b$ is proportional to the 
one-loop beta function coefficient 
\ref\RMat{M. Matone, \plt 357 (1995) 342.}%
\ref\RSTY{J. Sonnenschein, S. Theisen and  S. Yankielowicz,
\plt 367 (1996) 145.}.

From  $(A^\al)^*=B^\al$ one can derive the following useful relations 
satisfied by the functions $f_\al,\ c_\al$ and $h$ 
appearing in the ansatz for $\Pi$ and 
in $\kt$:
\eqn\sgid{\eqalign{
h=&\ -2iA\, u\, (s-\fc{1}{\gamma})+\mm\, u \, (\hx c_1\, s+ \hx c_2),
\ \  \hx c_i \in \bx C,\cr
0=&\ \sum_\al f_\al^2\eta_\al\cr
B=&\ \eta_\al f_\al c_\al= -\fc{{\ca}}{\bb {\ca}} b\cr
|A^2|=&\ |B^2|
}}
and
\eqn\noneos{
P_\al f_\al = 0 \ \ \Rightarrow  \ \ P_\al c_\al = 0.
}

\appendix{C}{The scalar potential for the universal volume modulus}
The
$\eps$ expansion of the matrices $V$ and $\tx U=U-V$ for the
section $\Pi$ in \falfs\ is given by
\eqn\uvmats{\eqalign{
\underline{\cx O(\eps^0):}\ \  &
e^{-K}\, V^{\al\bb \be}_0=f_\al \bb f_\be \qquad 
e^{-K}\, U_0^{\al\bb\be}=w_{\al,\rho}g^{\rho \bb \rho}\bb 
w_{\bb \be, \bb \rho}, \qquad
e^{-K}\, U_0^{l\bb k}=\fc{\cx V}{2}\ \tau^{l\bb k}
\qquad \qquad\qquad \qquad \qquad\qquad \qquad \qquad\ \cr\cr
\underline{\cx O(\eps^1):}\ \  
&e^{-K}\, V^{\al \bb k}_1=f_\al\, \bb {\ca}\bb a^{\bb k},\cr\cr
&e^{-K}\, \tx U^{\al \bb k}_1= \fc{V}{2|{\ca}|^2}\, \tau^{\bb k l}\,
\{A^\al u_l +\cc \},\cr\cr
\underline{\cx O(\eps^2):}\ \  
&e^{-K}\, V^{\al\bb \be}_2=
c_\al\bb f_\be\, u + \bb c_\be  f_\al\, \bb u ,\qquad
e^{-K}\, 
V_2^{k\bb k}=|{\ca}|^2\, a^k \bb a^{\bb k},\cr\cr
&
e^{-K}\, \tx U_2^{\al\bb\be}=
\fc{V}{2|{\ca}|^2}\, X^\al_k \tau^{k \bb l} \bb X^{\bb \be}_{\bb l}+
\{\fc{i b \Sigma}{2} (\eta_\al-\fc{f_\al}{\mm})\, \bb f_{\bb \be}\, u_s +\hc\}
\ + \cr
&\ \ \ \ \ \ \ \ \ \  \ \ \ \ \ \ \
\{-\fc{1}{\sqrt{2\mm}}\, 
(\kt_{\bb s}+\fc{2i}{\Sigma}\kt)\, 
f_\al\bb w_{\bb \be,\bb t} +\hc \}
\ +\cr
&\ \ \ \ \ \ \ \ \ \ \ \ \ \ \ \ \ 
\{i\sqrt{2\mm}\, 
(\eta_\al\, b u+\fc{\kt}{\mm\Sigma}f_\al)\, \bb w_{\bb \be,\bb t}+\hc\}\ 
-\ \fc{3\kt}{\Sigma}\, w_{\al,t}\, \bb w_{\bb \be,\bb t}
,
}}
where $t_\rho=\{t,s\}$ and
\eqn\exdefs{\eqalign{
A^\al&=-\fc{{\ca} b}{\mm}\{
\bb f_\bal+\fc{2\sqrt{2\mm}}{\gamma\, \Sigma}
\bb w_{\bal,\bb t} \},\cr
X^\al_k&=\fc{1}{\bb {\ca}}\, (A^\al\,  u_k +\cc)+
\fc{2\sqrt{2}i|{\ca}|^2}{\Sigma\sqrt{\mm}}\,
w_{\al,t}\, \tau_{k\bb l}\, \bb a^{\bb l},\cr
\tau_{k,\bb l}&=\Im a^k_{D,l}.
}}

\listrefs
\end

\end